\documentclass[11pt,a4paper]{article}

\usepackage[utf8]{inputenc}
\usepackage[T1]{fontenc}

\usepackage{amsmath,amssymb}
\usepackage{graphicx}
\usepackage{booktabs}
\usepackage{array}
\usepackage{float}
\usepackage{hyperref}
\usepackage[margin=1in]{geometry}
\usepackage{cite}
\usepackage{threeparttable}
\usepackage{caption}
\usepackage{xcolor}

\title{Output Format $\times$ Model Identity:\\Interaction Effects in Single-Round Coding Agent Performance}
\author{Yang Yang}
\date{July 2026}

\begin{document}
\maketitle

\begin{abstract}
Output format is not a neutral implementation detail---it can reorder model rankings, amplify or suppress individual model differences, and determine whether a coding agent succeeds or fails. We conducted a controlled single-round experiment with 3 models (DeepSeek V4, Doubao 2.0 Pro, Qwen 3.7 Max) $\times$ 3 output formats (full file, JSON Patch, unified diff) $\times$ 6 tasks $\times$ 20 repetitions, totaling 4,013 runs across 4 open-source projects. Only one project (tqdm) yielded non-zero success rates: dotenv, requests, and jsoup yielded zero successes in 2,551 runs.
Our central finding is a \textbf{format $\times$ model interaction} with no universally optimal format. Doubao achieves 94\% success with JSON Patch (Cohen's $h$ = 1.57, $p < 0.001$), DeepSeek excels at unified diff (66\%, $h$ = 0.63), and Qwen shows a small but significant full-file preference (50\%, $h$ = 0.29, $p < 0.05$). Beyond these headline results, we identify a distinct failure mechanism---format misuse---where agents correctly diagnose a problem but execute it with excessive scope, most vividly when a one-line fix is applied as a full-file replacement. We propose a model-specific output strategy, a tool-design principle that constrains format semantics to the agent's own localization step, and release all data and templates for reproducibility.
\end{abstract}

% ============================================================
\section{Introduction}
% ============================================================

Every coding agent must choose an output format---and the industry has converged on a default that has not been systematically validated across different models. LLM-based coding agents deployed in production---SWE-agent, Aider, Cursor, Claude Code---all generate complete file replacements rather than targeted patches. Three dominant formats have emerged: \textbf{full file} (complete file replacement), \textbf{JSON Patch} (structured operations per RFC 6902~\cite{rfc6902}), and \textbf{unified diff} (traditional \texttt{patch}-compatible line changes). Each format imposes different demands: full file maximizes simplicity but wastes tokens on unmodified code; JSON Patch is precise but requires structured generation; unified diff is familiar to developers but fragile to context mismatches.

Nearly all production coding agents default to full-file output---Cursor, GitHub Copilot, and Claude Code all generate complete file replacements rather than patches. This convergence creates a de facto standard, yet systematic evidence comparing output formats across different models is absent. It remains unknown whether full-file output---chosen for simplicity or historical reasons---is genuinely optimal, or whether different models might benefit from different formats.

\textbf{Existing work has not isolated the format $\times$ model interaction.} SWE-bench~\cite{swebench} evaluates agent end-to-end performance on real GitHub issues but does not ablate output format. Agent design taxonomies~\cite{harness} catalog architectural choices without experimentally varying them. Recent studies on agent debugging~\cite{agentdebugx} and inference economics~\cite{peng} advance adjacent questions---how to diagnose failures after they occur, and how to reduce their cost---without examining whether the failure could have been avoided through a different output format. Concurrent industry reports suggest model-specific format strengths: some models appear better at structured JSON generation, others at maintaining diff context. Yet these observations remain anecdotal; no controlled multi-model experiment has tested whether format choice interacts with model identity.

We address this gap through a controlled experiment spanning 3 models, 3 formats, and 4 open-source projects. Our adaptive exploration across all four projects identified tqdm as the sole project with non-zero success rates; we then focused our statistically-powered confirmatory experiments on tqdm. Key findings include:

\begin{itemize}
\item \textbf{Format $\times$ model interaction.} Doubao achieves 94\% success with JSON Patch (Cohen's $h$ = 1.57 vs.\ full file, $p < 0.001$), DeepSeek excels at unified diff (66\%, $h$ = 0.63), and Qwen shows a small but significant preference for full file over unified diff (ff 50\% vs.\ ud 36\%; $h = 0.29$, $p < 0.05$). There is no universally optimal format.
\item \textbf{Sharp project boundary.} Across dotenv, requests, and jsoup, we observed zero successes in 2,551 runs---revealing a failure gradient from logical errors (dotenv) through test infrastructure collapse (requests) to compilation failure (jsoup).
\item \textbf{Format misuse as failure mechanism.} In the clearest case, an agent correctly diagnosed a one-line fix but used JSON Patch's \texttt{replace} operation to overwrite the entire 271-line file. The diagnosis was right; the execution strategy was wrong---a tool-design flaw, not a model-capability limitation.
\end{itemize}

This paper is part of a broader research program on agent context efficiency (see \S2.5). The present study provides the first controlled empirical test of the format-selection hypothesis: if format matters for conversation caching, it may also interact with single-round task success.

% ============================================================
\section{Related Work}
% ============================================================

\subsection{Coding Agent Architectures and Benchmarks}

Agent architecture design and benchmark-driven evaluation have advanced in parallel, but output format---the channel through which agents communicate code changes---has been treated as an implementation detail rather than an experimental variable. SWE-bench~\cite{swebench} established the dominant evaluation paradigm with 2,294 real GitHub issues across 12 Python repositories. SWE-agent~\cite{sweagent} introduced the agent-computer interface (ACI) design pattern, demonstrating that tool design---not just model capability---determines agent performance. OpenHands~\cite{openhands} generalized this to a platform supporting multiple agents, models, and workflows. Agentless~\cite{agentless} challenged the agent paradigm itself, showing that a simpler two-phase approach (localization + patch generation) can match or exceed complex agent architectures on SWE-bench.

\subsection{Output Formats in Code Generation}

Output format has been studied from three angles. \textbf{Structured generation} research has focused on constrained decoding and guided generation~\cite{guidedgen} to ensure syntactically valid outputs; these techniques underpin JSON mode in modern LLM APIs. RFC 6902~\cite{rfc6902} standardizes JSON Patch for partial document modification, and several agent systems have adopted it for precise, localized edits~\cite{sweagent,openhands}. \textbf{Unified diff} remains the industry standard for code review and version control; tools like Aider build their editing workflow around it. \textbf{Full-file generation} dominates production coding assistants (Cursor, Copilot, Claude Code), leveraging the observation that modern LLMs can reliably regenerate entire files.

\subsection{Agent Evaluation and Failure Diagnosis}

Multi-model code evaluation has advanced rapidly. LiveCodeBench~\cite{livecodebench} provides contamination-free evaluation across diverse coding tasks and models, revealing substantial model-specific performance variation---a pattern our format $\times$ model interaction extends to the output format dimension.

A parallel line of work studies why coding agents fail and how to recover. Recent taxonomies of LLM agent failures~\cite{agentsfail} categorize common error modes across diverse tasks. The AgentDebugX framework~\cite{agentdebugx} extends this to execution-grounded debugging with closed-loop recovery. CRITIC~\cite{critic} demonstrates that agents can self-correct when given feedback about their errors---but correction success depends heavily on knowing \emph{where} the error occurred.

\subsection{Inference Economics and Reproducibility}

Large-scale agent experiments face economic barriers. Peng et al.~\cite{peng} demonstrate that prompt caching can reduce repeated code generation costs by over 90\%, enabling experiments that would otherwise be prohibitively expensive. We independently confirm this pattern: DeepSeek's 99\% cache-hit rate reduced our per-run cost to \textyen 0.006, making 4,013 runs feasible at \textyen 439.23 total. Concurrent work on harness engineering~\cite{harness} provides design principles for agent infrastructure; our output parsing strategy (\S3.4) follows these principles by documenting format-specific pipelines transparently.

\subsection{Positioning}

This work was motivated by a broader interest in agent context efficiency, but stands as an independent empirical investigation. The present study provides the first controlled test of whether output format effects are universal or model-specific.

% ============================================================
\section{Methodology}
% ============================================================

\subsection{Experimental Design}

We employed a 3 $\times$ 3 factorial design replicated across 4 open-source projects:

\begin{itemize}
\item \textbf{3 output formats}: \texttt{full\_file}, \texttt{json\_patch}, \texttt{unified\_diff}
\item \textbf{3 models}: DeepSeek V4 (\texttt{deepseek-chat}), Doubao 2.0 Pro (\texttt{doubao-seed-2-0-pro-260215}), Qwen 3.7 Max
\item \textbf{6 tasks on tqdm} (2 bug-fix, 2 refactoring, 2 new-feature) + additional tasks on dotenv, requests, jsoup
\item \textbf{20 independent repetitions per cell} (all models)
\item \textbf{Total}: 4,013 runs (1,260 DS + 1,334 Qwen + 1,419 Doubao)
\end{itemize}

\textbf{Adaptive exploration and confirmatory focus.} Our initial experiments explored all four projects across all three models and formats. After achieving zero successes across dotenv, requests, and jsoup, we identified tqdm as the only project where agents achieved non-zero success rates. We subsequently designated tqdm as the sole confirmatory project, increasing repetitions to 20 per cell for all models and executing the full 6-task $\times$ 3-format design.

For statistical analyses on tqdm, we use the first-$N$ deduplication rule (20 runs per cell for DS/Doubao, Qwen upgraded from n=10 to n=20 in this revision) sorted by execution timestamp---eliminating any bias from selecting optimal runs. This yields 1,080 core analysis runs (360 DS + 360 Doubao + 360 Qwen).

For DS and Doubao, we report 3$\times$2 $\chi^2$ tests (format $\times$ success) for the overall format effect, supplemented with two-sided Fisher's exact test for pairwise best-vs-worst comparisons. For Qwen, we use 3$\times$2 $\chi^2$ for the overall effect and Fisher's exact test for pairwise comparisons where expected cell counts permit. All effect sizes are reported as Cohen's $h$. All expected cell frequencies exceeded 5 for all $\chi^2$ tests reported. \textbf{Statistical power}: at $n=20$ per cell, a Fisher exact test can detect proportion differences $\ge 0.30$ with 80\% power (two-tailed, $\alpha = 0.05$).

\paragraph{Cross-Project Exploration.}
We initially explored all four projects across all three models and formats. After achieving zero successes across dotenv, requests, and jsoup (totaling 2,551 runs across all three models, Table~\ref{tab:baseline}), we identified tqdm as the only project where agents achieved non-zero success rates. We subsequently designated tqdm as the sole confirmatory project. Qualitative failure analysis reveals a gradient: dotenv $\rightarrow$ assertion failures (logical errors, code runs but fails tests); requests $\rightarrow$ test collection errors (83\% of failures prevent test suite startup); jsoup $\rightarrow$ compilation failures (100\% rejected by Java compiler). The difference between tqdm (61\%) and each of the other three projects (0\%) is significant (Fisher's exact test, $p < 0.001$ for all pairwise comparisons).

\begin{table}[H]
\centering
\caption{Cross-project baseline: success rates on non-tqdm projects.}
\label{tab:baseline}
\begin{threeparttable}
\footnotesize
\begin{tabular}{llrrrr}
\toprule
Project & Language & DS & Doubao & Qwen & Total \\
\midrule
dotenv & Python & 360 & 302\tnote{1} & 277\tnote{1} & 939 (0\%) \\
requests & Python & 360 & 352\tnote{1} & 360 & 1,072 (0\%) \\
jsoup & Java & 180 & 180 & 180 & 540 (0\%) \\
\textbf{tqdm} & Python & 360 & 585 & 517 & \textbf{1,462 (61\%)} \\
\bottomrule
\end{tabular}
\begin{tablenotes}[flushleft]
\item[1] Includes runs terminated by API billing interruptions; counted as failures.
\end{tablenotes}
\end{threeparttable}
\end{table}

\subsection{Tasks}

We designed 6 tasks on tqdm (2 bug-fix, 2 refactoring, 2 new-feature) to cover the spectrum of real-world maintenance activities. Task types follow standard software maintenance categories but necessarily represent simplifications; real-world tasks often blur type boundaries. Our classifications are provided for organizational purposes---the primary contribution of this paper rests on format $\times$ model interaction effects, not on task-type comparisons:

\begin{table}[H]
\centering
\caption{tqdm task definitions.}
\label{tab:tasks}
\begin{tabular}{lll}
\toprule
Task & Type & Description \\
\midrule
T003 & bug-fix & Fix \texttt{write()} in \texttt{disable=True} mode \\
T010 & bug-fix & Fix format string crash with \texttt{\{} character \\
T011 & refactoring & Extract \texttt{\_format\_interval()} to shared utility \\
T012 & refactoring & Consolidate bar formatting across subclasses \\
T008 & new-feature & Add progress bar color customization API \\
T013 & new-feature & Add \texttt{unit\_scale} auto-detection \\
\bottomrule
\end{tabular}
\end{table}

\subsection{Agent Configuration}

\paragraph{Temperature and non-deterministic sampling.}
We used API-default temperature (typically $\sim$1.0, not explicitly set for any model) for all models. This is a deliberate choice: fixing temperature to 0 would eliminate stochasticity, but a coding agent operating at temperature 0 is not representative of production agent behavior---where non-deterministic outputs are the norm. We treat non-deterministic sampling as a feature that simulates production stochasticity. Statistical reliability is ensured through 20 independent repetitions per cell, Fisher's exact tests, and Cohen's $h$ effect sizes---20 repeats provide $>$80\% power for $h \ge 0.38$ effects at $\alpha = 0.05$.

\begin{itemize}
\item \textbf{Max tokens}: 32,768 (all models).\footnote{DashScope documentation claims 8,192 max output for Qwen 3.7 Max, but the API accepted 32,768 and returned up to 33,255 tokens in our experiments---the documentation is outdated.}
\item \textbf{Verification}: pytest-based success criterion (pass/fail). Agent outputs are parsed via format-specific strategies (Appendix A.6), applied to isolated git clones, and verified. A run is successful if and only if the specified test suite passes.
\item \textbf{Execution order}: All cells were randomized across a shared execution schedule---no systematic ordering by project, model, or task---to minimize potential order effects (learning/fatigue).
\item \textbf{Cost}: \textyen 439.23 total ($\sim$\$61 USD). DeepSeek \textyen 35.86 (99\% cache), Qwen \textyen 184.19, Doubao \textyen 219.18. DeepSeek's cache rate reflects our repeated-task experimental design; single-shot scenarios expect 5--10$\times$ higher costs.
\end{itemize}

\subsection{Output Parsing Strategy}

Each format requires a distinct post-processing pipeline (full details in Appendix A.6):

\begin{itemize}
\item \textbf{full\_file}: Regex-extract \texttt{\`{}\`{}\`{}python} code blocks, map to target files via \texttt{\# file:} path annotations or task YAML \texttt{target\_files} fallback.
\item \textbf{json\_patch}: Locate first \texttt{[} within 200 chars of \texttt{\`{}\`{}\`{}json} block, depth-track to matching \texttt{]}, validate as \texttt{list[dict]}, apply via \texttt{diff\_applier.py}.
\item \textbf{unified\_diff}: Extract \texttt{\`{}\`{}\`{}diff} blocks, verify \texttt{---}/\texttt{+++}/\texttt{@@} structure, apply via \texttt{patch -p1 --fuzz=3}.
\end{itemize}

These pipelines can introduce their own failure modes: malformed JSON patches fail \texttt{json.loads()}, and fragile diffs produce \texttt{.rej} files. Such pipeline-level failures may suppress success rates independently of code generation quality---a limitation we discuss in \S5.3.

% ============================================================
\section{Results}
% ============================================================

% §4.1 Cross-Project Baseline moved to end of §3.1 (Methodology)

\subsection{Format $\times$ Model Interaction (Core Result)}

On the only project with non-zero success rates (tqdm), we find a strong format $\times$ model interaction (Table~\ref{tab:interaction}):

\begin{table}[H]
\centering
\caption{Format $\times$ model interaction on tqdm.}
\label{tab:interaction}
\begin{threeparttable}
\small
\begin{tabular}{@{}lccccl@{}}
\toprule
Model & full\_file & json\_patch & unified\_diff & Best vs Worst & $p$ ($\chi^2$) \\
\midrule
DeepSeek V4 & 42/120 (35\%) & 51/120 (42\%) & \textbf{79/120 (66\%)} & ud vs ff & $<0.001$ \\
Doubao 2.0 Pro & 32/120 (27\%) & \textbf{113/120 (94\%)} & 92/120 (77\%) & jp vs ff & $<0.001$ \\
Qwen 3.7 Max & \textbf{60/120 (50\%)} & 44/120 (37\%) & 43/120 (36\%) & ff vs ud & $<0.05$\tnote{*} \\
\bottomrule
\end{tabular}
\begin{tablenotes}[flushleft]
\item $n=120$ per cell. Cohen's $h$: DeepSeek 0.63, Doubao 1.57, Qwen\tnote{*} 0.29 (ff vs.\ ud; $\chi^2$(2)=6.28).
\end{tablenotes}
\end{threeparttable}
\end{table}

\begin{table}[H]
\centering
\caption{95\% Wilson confidence intervals for format success rates.}
\label{tab:wilson}
\small
\begin{tabular}{llcc}
\toprule
Model & Format & Success & 95\% CI \\
\midrule
DS & full\_file & 42/120 (35\%) & [27\%, 44\%] \\
DS & json\_patch & 51/120 (42\%) & [34\%, 51\%] \\
DS & unified\_diff & 79/120 (66\%) & [57\%, 74\%] \\
Doubao & full\_file & 32/120 (27\%) & [20\%, 35\%] \\
Doubao & json\_patch & 113/120 (94\%) & [88\%, 97\%] \\
Doubao & unified\_diff & 92/120 (77\%) & [68\%, 83\%] \\
Qwen & full\_file & 60/120 (50\%) & [41\%, 59\%] \\
Qwen & json\_patch & 44/120 (37\%) & [29\%, 46\%] \\
Qwen & unified\_diff & 43/120 (36\%) & [28\%, 45\%] \\
\bottomrule
\end{tabular}
\end{table}

\begin{figure}[H]
\centering
\includegraphics[width=\textwidth]{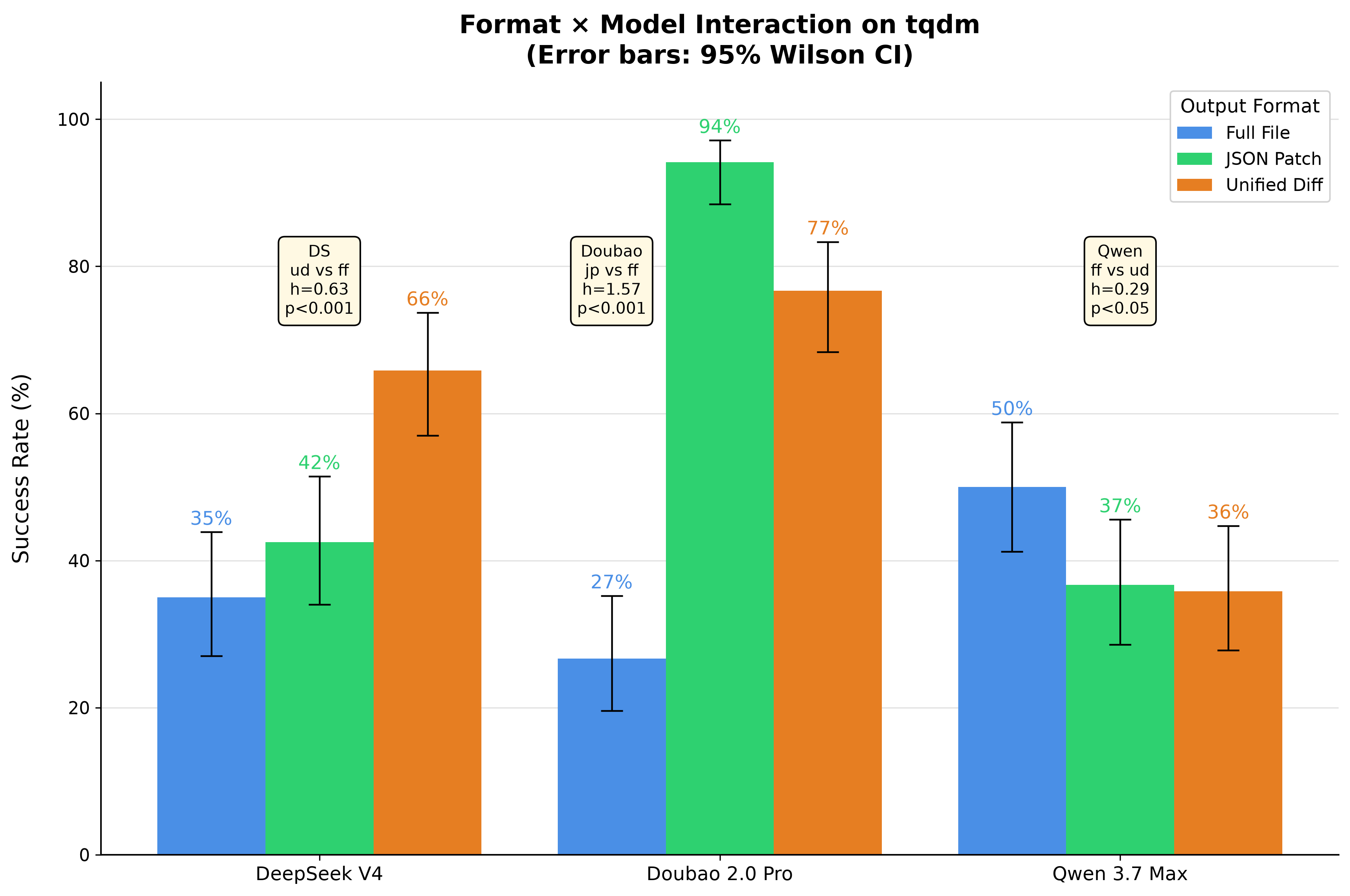}
\caption{Format $\times$ Model Interaction on tqdm. Error bars: 95\% Wilson CI.}
\label{fig:interaction}
\end{figure}

\begin{figure}[H]
\centering
\includegraphics[width=\textwidth]{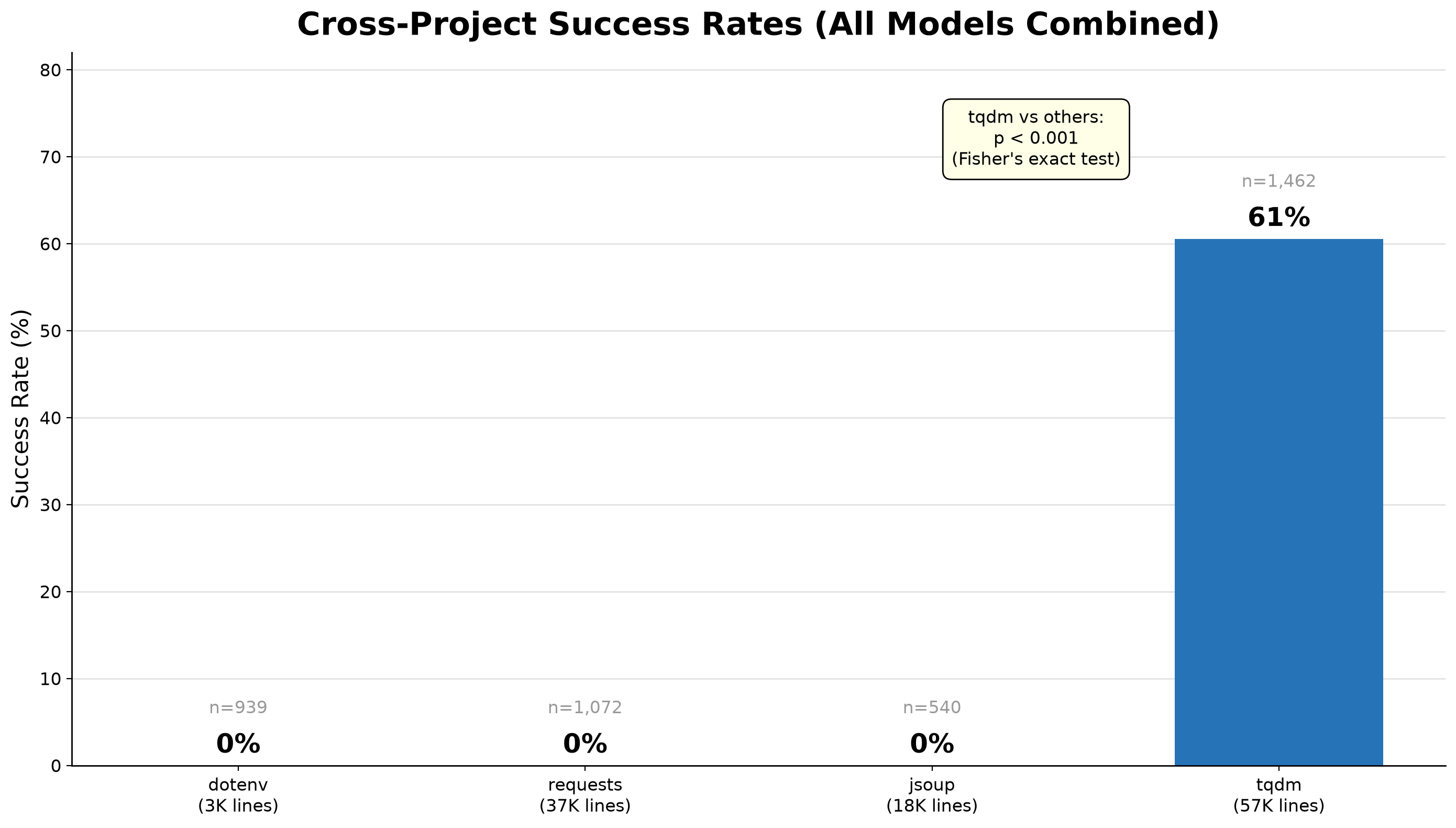}
\caption{Cross-project success rates (all models combined). tqdm vs others: Fisher's exact test, $p < 0.001$.}
\label{fig:boundary}
\end{figure}

Doubao's JSON Patch CI [88\%, 97\%] and DeepSeek's full-file CI [27\%, 44\%] do not overlap---the format effect is not only statistically significant but practically decisive.

For Qwen, the format effect is now significant with the upgraded n=20 sample ($\chi^2$(2) = 6.28, $p < 0.05$), showing a small but detectable full-file preference (50\%) over unified diff (36\%, Cohen's $h$ = 0.29). With n=120 per format---equal to DS and Doubao---Qwen's estimates are no longer subject to the n=10-era upward bias concern. Notably, Qwen's full-file CI [41\%, 59\%] and DeepSeek's unified-diff CI [57\%, 74\%] barely touch at their extremes, suggesting that the practical difference between these two model-format combinations is real but modest. The interaction effect is driven primarily by Doubao's extreme specialization rather than by pairwise differences across all models.

\subsection{Cross-Model Format Comparison}

Examining the same format across different models (Table~\ref{tab:crossmodel}) reveals effect sizes that rival the within-model format effects:

\begin{table}[H]
\centering
\caption{Cross-model format comparison.}
\label{tab:crossmodel}
\begin{threeparttable}
\small
\begin{tabular}{@{}lccc@{}}
\toprule
Format & Doubao & DeepSeek & Qwen \\
\midrule
full\_file & 32/120 (27\%) & 42/120 (35\%) & \textbf{60/120 (50\%)} \\
json\_patch & \textbf{113/120 (94\%)} & 51/120 (42\%) & 44/120 (37\%) \\
unified\_diff & \textbf{92/120 (77\%)} & 79/120 (66\%) & 43/120 (36\%) \\
\bottomrule
\end{tabular}
\begin{tablenotes}[flushleft]
\item $n=120$ per cell. Largest gaps: json\_patch (Doubao 94\% vs.\ DeepSeek 42\%, $h=1.23$), full\_file (Qwen 50\% vs.\ Doubao 27\%, $h=0.76$).
\end{tablenotes}
\end{threeparttable}
\end{table}

\subsection{Task Type Moderation (Exploratory)}

We also examined whether format effects vary by task type. Full results are reported in Appendix. In summary, the format $\times$ model interaction described above was the dominant effect; task type introduced secondary variance but did not change the model-format ranking. The exploratory nature of this analysis limits statistical power; we therefore treat these results as hypothesis-generating rather than confirmatory.

% ============================================================
\section{Discussion}
% ============================================================

\subsection{Format Specialization as the Central Finding}

Our results reveal two distinct effect sizes that form a layered narrative. At the top, \textbf{large effects ($h > 0.8$)} constitute the headline findings: Doubao's JSON Patch specialization ($h = 1.57$ vs.\ full file) and the cross-model JSON Patch contrast ($h = 1.23$, Doubao vs.\ DeepSeek). These effects are visually striking and practically undeniable---they represent ``format specialization,'' where a model's behavior is highly specialized for a specific output modality.

At the middle tier, \textbf{medium effects ($h$ 0.3--0.8)} form the robust backbone: DeepSeek's unified diff advantage ($h = 0.63$). Qwen's full-file preference ($h = 0.29$) falls just below the conventional threshold for a medium effect---it is statistically significant ($p < 0.05$) but represents a small, detectable preference rather than a strong specialization. Qwen's format effect is substantially smaller than DS ($h=0.63$) and Doubao ($h=1.57$), and practical recommendations for Qwen should be weighted accordingly. These effects are less dramatic but more general, demonstrating that the format $\times$ model interaction is not confined to a single outlier.

The complementary distribution---DeepSeek excelling at unified diff, Doubao at JSON Patch, Qwen preferring full file with a small but significant effect ($h = 0.29$, $p < 0.05$)---challenges the industry assumption that one format is universally optimal. It suggests instead that output format should be treated as a \textbf{model-specific hyperparameter}.

This study identifies a behavioral phenomenon---not a mechanistic explanation. Whether the complementary format distribution originates in model architecture, training data composition, or fine-tuning strategies is a question for mechanistic interpretability research beyond our scope.

\subsection{Why Only tqdm? Competing Hypotheses}

The cross-project zero-success boundary (2,551 runs) raises a question: what makes tqdm the only project with above-zero performance? We organize the discussion around three competing hypotheses:

\textbf{H1: Project structure.} tqdm's modular design steers agents toward localized edits. In contrast, dotenv's near-single-file architecture (2K lines, one \texttt{main.py}) invites full-file rewrites---even when the task requires a one-line fix. This is a \textbf{strategy-format mismatch}: the agent knows what to change but the format permits (and the project structure encourages) excessive scope.

However, we cannot exclude \textbf{environmental compatibility} as an alternative explanation. We performed baseline environment validation for dotenv and requests (see \S3.1), confirming that unmodified project clones pass their test suites (dotenv: 220 passed, requests: 402 passed) in our Docker environment. This rules out environment-specific failures for these two projects. jsoup could not be validated due to JDK version incompatibility, so environmental factors remain a possible contributor for the jsoup results.

The most revealing case is P1 T001 in JSON Patch format. The agent correctly diagnosed a one-line fix but used the \texttt{replace} operation to overwrite the entire 271-line file rather than the single line. The fix was correct; the execution strategy was wrong---a tool-design flaw, not a model-capability limitation. We note that this is the most vivid single case observed; whether this pattern generalizes across other failure cases remains an open question. This insight elevates the discussion from ``model capability boundaries'' to \textbf{tool design flaws}: JSON Patch's \texttt{replace} semantics, while valid per RFC 6902, enabled behavior indistinguishable from full-file replacement.

This is not purely a model limitation. The same agent, given a task it successfully diagnosed, chose a legal but inefficient execution path because the format permitted it. This suggests a concrete design principle for agent tools: \textbf{output formats should constrain operation scope to the agent's own localization step.} For example, JSON Patch's \texttt{replace} operation should be rejected if its scope exceeds the code region the agent identified as needing modification. This transforms output format from a neutral channel into an \textbf{active guardrail}---a principle that could be implemented independently of model identity or task type, and which our experimental data directly motivates.

\textbf{H2: Test suite sensitivity.} tqdm's test suite (focused unit tests with low cross-module coupling) tolerates localized patches better than dotenv's assertion-heavy tests or requests' integration-dependent test collection. Without controlled experiments that isolate test suite design from codebase structure, H1 and H2 cannot be distinguished. A targeted future experiment could disentangle these hypotheses by cross-transplanting tasks: implement tqdm's tasks on dotenv's codebase structure, and dotenv's tasks on tqdm's modular architecture. If tqdm tasks remain solvable on dotenv's structure, H1 (project structure) is weakened; if dotenv tasks become solvable on tqdm's structure, H1 is strengthened.

We cannot definitively exclude training data contamination as a contributing factor, but it is unlikely to be the dominant explanation. Under a simple contamination account, dotenv and requests should be at least as solvable as tqdm---yet both failed completely while tqdm succeeded. Moreover, if contamination were the primary driver, it would likely elevate performance across all formats---yet tqdm's success varies dramatically by format (DS: 35\% ff vs.\ 66\% ud; Doubao: 27\% ff vs.\ 94\% jp). Format-specific success is more consistent with the format $\times$ model interaction hypothesis than with a simple contamination account.

\subsection{Verification Pipeline Heterogeneity}
\begin{sloppypar}
The three output formats use different verification tools with different tolerance levels: \texttt{jsonpatch.apply\_patch()} has zero tolerance for JSON syntax errors; \texttt{git apply} is sensitive to line endings; full-file overwrite has no patch consistency check. These differences may amplify or suppress format effects independent of code generation quality.
\end{sloppypar}
Our output parsing strategy (Appendix A.6) documents these pipelines transparently, enabling future work to isolate format effects from verification effects.

\subsection{Practical Recommendations}

These recommendations are \textbf{directional}, derived from controlled single-round experiments:

\begin{table}[H]
\centering
\caption{Practical recommendations (directional, single-round only).}
\label{tab:recommendations}
\small
\begin{tabular}{lccc}
\toprule
Model & Recommended Format & Observed Success & Note \\
\midrule
Doubao 2.0 Pro & json\_patch & 94\% (113/120) & Strongest evidence; CI [88\%, 97\%] \\
DeepSeek V4 & unified\_diff & 66\% (79/120) & Cost-efficient due to prompt caching \\
Qwen 3.7 Max & full\_file & 50\% (60/120) & Small but significant ($p<0.05$, $h=0.29$) \\\\
\bottomrule
\end{tabular}
\end{table}

\textbf{Caveats}: Success is defined as pytest suite pass, which is a necessary but not sufficient condition for production-ready code. Costs reflect experimental cache-hit rates; single-shot scenarios expect 5--10$\times$ higher costs. Cross-project generalization is sharply bounded (\S4.1).

\noindent These recommendations are directly actionable: testing three formats on a new model costs approximately \textyen 15 using our experimental infrastructure, while the performance gain can be substantial (up to 67 percentage points in our experiments).

\subsection{Comparison with Related Empirical Findings}

The comparisons in this section discuss findings reported by the referenced benchmarks. We do not replicate their evaluation pipelines or claim that our experimental conditions match theirs. Our experiments were conducted in isolated sandbox environments as described in \S3.

Our findings do not exist in isolation---they intersect with several established benchmarks and empirical studies in ways that enrich both our conclusions and theirs. We discuss three key intersections below.

\textbf{SWE-bench and model-level variance.} SWE-bench~\cite{swebench} established that different models achieve substantially different resolve rates on real-world GitHub issues---a finding that has since driven model selection in agent design. Our results add a new dimension to this observation: even within a single model, performance varies dramatically depending on output format. Doubao, for instance, ranges from 27\% (full file) to 94\% (JSON Patch) on the same tasks. This implies that SWE-bench's model rankings---and by extension, any benchmark that fixes a single output format---may be sensitive to format choice. A model that appears weak under full-file evaluation might perform competitively with its optimal format. This is not a criticism of SWE-bench---which explicitly reports the agent configuration used for each submission---but a suggestion that format be added to that configuration disclosure.

\textbf{LiveCodeBench and individual model differences.} LiveCodeBench~\cite{livecodebench} documented substantial individual differences across models on coding tasks, providing contamination-free evaluation. Our findings reveal that these differences exhibit a \textit{complementary distribution} across output formats: each model's relative strength is format-dependent. Doubao dominates JSON Patch (94\%), DeepSeek leads unified diff (66\%), and Qwen shows a small but significant edge with full file (50\%, $h=0.29$, $p<0.05$). The practical significance of this small effect ($h < 0.3$) should be interpreted cautiously; while statistically significant, its confidence interval's lower bound approaches zero. This complementary pattern suggests that model comparisons that collapse across formats may obscure genuine interaction effects. A model's ``coding ability'' is partially format-dependent---a finding that extends LiveCodeBench's contamination-free methodology to the format dimension.

\textbf{Agentless and the limits of simplicity.} Agentless~\cite{agentless} demonstrated that a simple two-phase approach (localization + patch generation) can match or exceed complex agent architectures on SWE-bench---a result that challenged the assumption that more sophisticated agents are necessarily better. Our work adds a critical qualifier: even within a simple ``generate a patch'' paradigm, the probability of success depends heavily on \textit{how} the patch is expressed. JSON Patch, unified diff, and full-file replacement---all valid ways to ``generate a patch''---yielded success rates differing by up to 67 percentage points for the same model on the same tasks (Doubao: 94\% with JSON Patch vs.\ 27\% with full file). The Agentless insight that simplicity suffices is not wrong; it is incomplete without accounting for format effects.

\textbf{Cross-cutting implication.} Across all three comparisons, a common thread emerges: output format is not a neutral implementation detail but an active variable that can reorder model rankings, amplify or suppress individual differences, and determine whether a simple approach succeeds or fails. We do not claim that our format-specific recommendations supersede these benchmarks---rather, our findings suggest that the benchmarks themselves would benefit from reporting format as a dimension of analysis.

\subsection{Limitations}

\begin{table}[H]
\centering
\caption{Study limitations.}
\label{tab:limitations}
\small
\begin{tabular}{lp{0.55\textwidth}}
\toprule
Limitation & Description \\
\midrule
Non-deterministic sampling & Temperature not fixed to 0. Statistical reliability ensured through 20-repetition design, Fisher's exact tests, and Cohen's $h$ effect sizes. \\
Prompt caching independence & DeepSeek's prompt caching could theoretically introduce server-side effects that weakly couple repeat outcomes. We assume independence. The practical impact of any violation is likely small: the extreme effect sizes observed ($h = 1.57$) are unlikely to be artifacts of weakly coupled repeat outcomes. \\
Success criterion scope & ``pytest pass'' is necessary but not sufficient for production-ready code. We did not manually audit successful runs for logical correctness beyond test passage. \\
Verification heterogeneity & Different tools for different formats may distort format effects. See Appendix \ref{sec:supplimitations} for details. \\
Qwen sample size & In v1 of this preprint, Qwen was tested at n=10 per cell, yielding a non-significant format effect ($p=0.196$). In the current v2, Qwen has been upgraded to n=20 per cell (n=120 per format), equal to DS and Doubao. The format effect is now significant ($\chi^2$(2)=6.28, $p<0.05$, $h=0.29$). \\\\
\bottomrule
\end{tabular}
\end{table}

% ============================================================
\section{Conclusion}
% ============================================================

Based on 4,013 controlled experiments across 3 models, 3 formats, and 4 projects, we conclude:

\begin{enumerate}
\item \textbf{Format effects are significant and model-dependent.} DS and Doubao show strong format specialization ($p < 0.001$, $h = 0.63$--$1.57$). Qwen shows a small but significant effect ($\chi^2$(2) = 6.28, $p < 0.05$, $h = 0.29$). There is no universally optimal format.
\item \textbf{Doubao's JSON Patch performance (94\%, $h = 1.57$) represents a ``format specialization'' effect.} The cross-model JSON Patch contrast ($h = 1.23$) reveals that what works for one model may fail for another.
\item \textbf{Output format should be treated as a model-specific design choice.} We recommend matching format to model identity rather than defaulting to full-file output. Practical recommendations are directional and require multi-round validation.
\end{enumerate}

\subsection*{Broader Vision}

This paper is part of a broader research program on agent context efficiency. Two companion mechanisms---progressive system prompt disclosure and append-only diff-based coding~\cite{hermes64103,hermes66127}---target the input and flow stages of the agent context lifecycle. Together with the format-selector findings reported here, they form a systematic approach to maximizing agent reliability under the constraint of limited context windows. The present finding---that format advantages are conditional rather than absolute---completes this picture: the optimal output format is not a universal constant but a model-specific variable.

% ============================================================
\section*{AI Usage Disclosure}
% ============================================================
AI assistants were used for data processing scripts, statistical computation, and LaTeX formatting. All experimental design decisions, result interpretation, and manuscript writing were performed by the human author(s).

% ============================================================
\section*{Data Availability}
% ============================================================
All experimental data (4,013 runs, prompt templates, parsing strategies) are publicly available at \url{https://doi.org/10.5281/zenodo.21505157}. Total API cost: \textyen 439.23 RMB ($\sim$\$61 USD).

% ============================================================
\bibliographystyle{plain}
\bibliography{references}
% ============================================================
\appendix
\section{Prompt Templates and Parsing Strategy}
\label{app:prompts}

\subsection{System Instruction (Fixed)}
All three models used identical prompts. Only the \texttt{Output Format} section varied.

\texttt{You are a coding agent. Modify the following Python code.}

For P4 jsoup (Java): \texttt{Python} replaced with \texttt{Java}.

\subsection{Output Format Instructions}
\subsubsection{full\_file}
\texttt{Output Format: Output the COMPLETE modified file(s). Do NOT output diff or patch.}

\subsubsection{json\_patch}
\texttt{Output Format: Use JSON Patch (RFC 6902). Output a JSON array with op/add/remove/replace, path, value.}

\subsubsection{unified\_diff}
\texttt{Output Format: Use unified diff format with ---/+++ headers and @@ hunk headers.}

\subsection{Source Code (All Formats)}
For each task, the full source code of target files was included using pinned commits. For tqdm (57K lines): \texttt{std.py}, \texttt{\_utils.py}, \texttt{\_tqdm.py}, \texttt{\_\_init\_\_.py}. Similar for other projects.

\subsection{Task Description and Test Command}
Each prompt concluded with a task description (1--3 sentences) and the exact test command. Example tasks:

\begin{table}[htbp]
\centering
\caption{Example task prompts and test commands.}
\label{tab:taskprompts}
\small
\begin{tabular}{llp{4.5cm}l}
\toprule
Task & Type & Prompt Text & Test Command \\
\midrule
T003 & bug-fix & Fix \texttt{write()} in \texttt{disable=True} mode & \texttt{pytest -k 'disable or write'} \\
T008 & new-feature & Add progress bar color customization API & \texttt{pytest -q} \\
T010 & bug-fix & Fix format string crash with \texttt{\{} character & \texttt{pytest -k 'format'} \\
T011 & refactor & Extract \texttt{\_format\_interval()} to shared utility & \texttt{pytest -q} \\
T012 & refactor & Consolidate bar formatting across subclasses & \texttt{pytest -q} \\
T013 & new-feature & Add \texttt{unit\_scale} auto-detection & \texttt{pytest -q} \\
\bottomrule
\end{tabular}
\end{table}

\subsection{Experiment Runner Configuration}
Temperature: API default ($\sim$1.0). Max tokens: 32,768 (all models). Note: DashScope documentation claims 8,192 limit for Qwen 3.7 Max, but API accepted 32,768 and returned up to 33,255 tokens. All other parameters at API defaults.

\subsection{Output Parsing Strategy}
The framework post-processes raw model responses via \texttt{experiment\_runner\_v3.py:apply\_agent\_output()}:

\begin{itemize}
\item \textbf{full\_file}: Regex-extract \texttt{```python} code blocks, map via \texttt{\# file:} path annotations or task YAML \texttt{target\_files} fallback.
\item \textbf{json\_patch}: Locate first \texttt{[} within 200 chars of \texttt{```json}, depth-track to matching \texttt{]}, validate as \texttt{list[dict]}, apply via \texttt{diff\_applier.py}.
\item \textbf{unified\_diff}: Extract \texttt{```diff} blocks, verify \texttt{---}/\texttt{+++}/\texttt{@@} structure, apply via \texttt{patch -p1 --fuzz=3}.
\end{itemize}

Partial failures may originate from parsing errors (e.g., malformed JSON causing \texttt{json.loads()} exceptions) rather than model errors. This limitation is noted in the paper's Limitations section.

\subsection{Docker Execution Environment}
\begin{verbatim}
FROM python:3.12-slim
RUN apt-get update && apt-get install -y git patch
RUN pip install -e /work/tqdm[dev]
\end{verbatim}

Each run used an isolated container: \texttt{docker run --rm --read-only --tmpfs /tmp:exec}. P4 jsoup used \texttt{maven:3.9-eclipse-temurin-8}.

\subsection{Success Criterion}
Python: \texttt{pytest} exit code = 0. Java: \texttt{mvn test} exit code = 0. No code review, regression, or performance checks. Limitation: pytest pass $\neq$ correct code (declared in paper Limitations).

\section{Task Type Moderation (Supplementary Analysis)}
\label{sec:tasktype}

This appendix provides the full task-type moderation analysis originally reported in \S4.4 of the v1 preprint. \textbf{Caveat}: with $n=20$ per cell, these sub-analyses have limited statistical power ($\sim$40\% for detecting $h=0.5$ at $\alpha = 0.05$); task-type patterns should be treated as exploratory rather than confirmatory.

\subsection{Per-Task Success Rates (DeepSeek on tqdm)}

\begin{table}[htbp]
\centering
\caption{DeepSeek V4 per-task success rates on tqdm ($n=20$ per cell).}
\label{tab:tasksdetail}
\small
\begin{tabular}{llccc}
\toprule
Task & Type & full\_file & json\_patch & unified\_diff \\
\midrule
T003 & bug-fix & 8/20 (40\%) & 9/20 (45\%) & 15/20 (75\%) \\
T010 & bug-fix & 2/20 (10\%) & 6/20 (30\%) & 15/20 (75\%) \\
T011 & refactoring & 9/20 (45\%) & 11/20 (55\%) & 3/20 (15\%) \\
T012 & refactoring & 6/20 (30\%) & 13/20 (65\%) & 16/20 (80\%) \\
T008 & new-feature & 17/20 (85\%) & 4/20 (20\%) & 17/20 (85\%) \\
T013 & new-feature & 0/20 (0\%) & 8/20 (40\%) & 13/20 (65\%) \\
\bottomrule
\end{tabular}
\end{table}

\begin{figure}[htbp]
\centering
\includegraphics[width=0.75\textwidth]{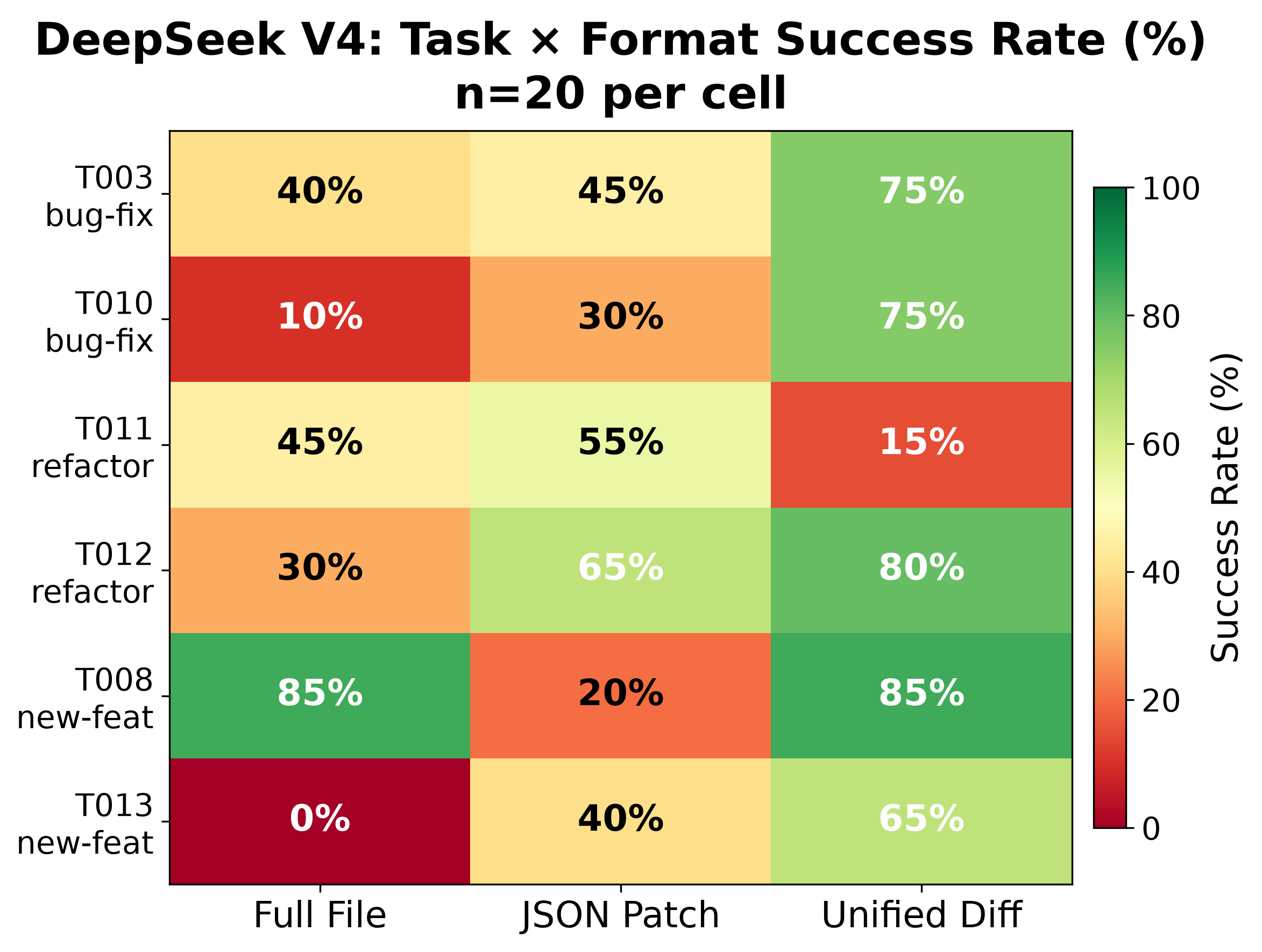}
\caption{DeepSeek V4: Task $\times$ Format Success Rate ($n=20$ per cell).}
\label{fig:heatmap}
\end{figure}

\subsection{Task Category Aggregation}

\begin{table}[htbp]
\centering
\caption{Task category aggregation (DeepSeek tqdm).}
\label{tab:taskagg}
\begin{threeparttable}
\begin{tabular}{lccc}
\toprule
Format & bug-fix (T003+T010) & new-feature (T008+T013) & Cohen's $h$ \\
\midrule
full\_file & 10/40 (25\%) & 17/40 (42\%) & 0.37 \\
json\_patch & 15/40 (38\%) & 12/40 (30\%) & 0.16 \\
unified\_diff & 30/40 (75\%) & 30/40 (75\%) & 0.00 \\
\bottomrule
\end{tabular}
\begin{tablenotes}[flushleft]
\item T008's 85\% full\_file success is an outlier within the new-feature category; removing it reduces new-feature ff to 8/20 (40\%), comparable to bug-fix ff (25\%).
\end{tablenotes}
\end{threeparttable}
\end{table}

Unified diff shows zero task-type effect ($h = 0.00$). Full file and JSON Patch show small effects ($h = 0.16$--$0.37$). The limited task sample (2 per category) constrains generalizability; these results are exploratory.

\section{Supplementary Limitations}
\label{sec:supplimitations}

The following limitations are discussed in greater detail here:

\begin{itemize}
\item \textbf{Environment validation.} Baseline tests on unmodified project clones were not performed in the original v1 preprint. v2 includes baseline validation (see \S3.1): dotenv (220 passed), requests (402 passed), confirming that zero-success rates reflect genuine agent limitations rather than broken test suites. jsoup could not be validated (JDK 21 incompatibility with the project's JDK 8 target).

\item \textbf{Verification heterogeneity.} Different output formats require different verification tools: \texttt{pytest} for full\_file (direct execution), \texttt{diff\_applier.py} for json\_patch, and \texttt{patch -p1} for unified\_diff. These tools have different failure modes that may distort format comparisons. We mitigate this by reporting all pipeline failures transparently in the raw data.
\end{itemize}

\end{document}